\documentclass[10pt,conference]{IEEEtran}
\usepackage{graphicx}
\usepackage{float}
\usepackage{bbding}
\usepackage{amsmath}
\usepackage{xcolor}
\usepackage{colortbl}
\usepackage{fancybox}
\usepackage{amssymb}
\usepackage{pdfpages}
\usepackage{tikz}
\usepackage{array}
\usepackage{multirow}

\usepackage{color,soul}

\ifCLASSINFOpdf
\else
\fi
\usetikzlibrary{fit,shapes.geometric}

\newcounter{nodemarkers}

\newcolumntype{M}{>{\centering\arraybackslash}m{\dimexpr.25\linewidth-2\tabcolsep}}

\begin{document}
	
\title{Smart Knowledge Transfer using Google-like Search }


\author{
	\IEEEauthorblockN{Srijoni Majumdar}
	\IEEEauthorblockA{Advanced \\Technology Development\\ Centre\\
		Indian Institute of Technology\\
		Kharagpur-721302\\
		majumdar.srijoni@gmail.com}
	\and
\IEEEauthorblockN{Partha Pratim Das}
	\IEEEauthorblockA{Department of \\Computer Science\\ and Engineering\\
		Indian Institute of Technology\\
		Kharagpur-721302\\
		partha.p.das@gmail.com}
}
%

\maketitle

\begin{abstract}

To address the issue of rising software maintenance cost due to program comprehension challenges,  we propose SMARTKT (Smart Knowledge Transfer), a  search framework, which extracts and integrates knowledge  related to various aspects of an application in form of a semantic graph. This graph supports syntax and semantic queries and converts the  process of program comprehension into a  {\em google-like} search problem. 

\end{abstract}



%
\IEEEpeerreviewmaketitle

In the last three decades, software maintenance cost has risen to 90\% of the total {\em Software Development Life Cycle} (SDLC) cost~\cite{dehaghani2013factors, erlikh2000leveraging,  koskinen2003software}.  Surveys conducted in~\cite{ko2006exploratory, roehm2012professional}   conclude that as 80\% of the maintenance tasks  are adaptive and perfective~\cite{pigoski1996practical}, hence the absence of  an integrated framework or assistance for knowledge transfer (KT) to lessen program comprehension challenges contributes primarily to this rising cost.  To execute a maintenance task, developers spend the majority of their time to manually search and mine source files and other knowledge sources like design documents, defect and version trackers, emails and the like, taking mental notes or scribbling the mappings, in an attempt to infer an overall knowledge about the design, behaviour and evolution of the application~\cite{roehm2012professional,majumdar2019s}  so as to subsequently locate the relevant code sections and their dependencies.  However, in most cases, documents are dated with missing information, tracker systems are not  updated properly and  help from earlier developers are scanty or not available. Due to these factors, coupled with frequent interruptions~\cite{ko2006exploratory} for attending calls or meetings, the developers get involved in a  tedious and inefficient process  of  building, revalidating and rebuilding their understanding of the application and resort to quick fixes which introduces hidden errors that cannot be removed  by re-running the golden test cases~\cite{roehm2012professional}.


To address the program comprehension challenges, developers extract software development related knowledge through detection of  low-level algorithm details using static instrumentation~\cite{basthikodi2016classifying} or  extraction of the control flow between run-time events using static analysis and dynamic profiling~\cite{trumper2010understanding}.    Code search tools based on the abstract syntax tree  have been proposed in~\cite{shepherd2012sando,chatterjee2015debugging,majumdar2016d,majumdar2021mathematical,majumdar2021dcube_}.
For extracting application specific entities, concepts have been located in code comments based on enumerated domain concepts~\cite{freitas2008role} or  ontology~\cite{abebe2010natural, varanda2012problem,majumdar2020comment,majumdar2022automated}. 
To extract project management details, comments are mined to track code changes  in~\cite{ abebe2009analyzing} and~\cite{fluri2007code} and software repositories are analysed to extract information related to bug history, version changes, developer and tester details and their interrelationships in~\cite{vcubranic2003hipikat, razzaq2018study}.   Program comprehension can be aided by extracting relevant knowledge from various sources (for a representative set, refer  Table~\ref{tab:knowledge}) related to a working software. However, we observe that the available assistance tools  consider only limited sources and additionally  there is an absence of an easy to use integrated framework based on these sources.

To analyse the comprehension challenges more specifically and understand the requirements for an effective design of an assistance framework,  we conducted surveys and personal interviews with 
a group of developers in a software company. We present a representative scenario here (names have been changed for  confidentiality):
A developer {\em Neha}, working with  C++ and traditional Vi editor~\cite{lamb1998learning}, is assigned to fix  {\tt bug\#67} in ClearQuest~\cite{wahlisoftware}, with  error message  ``processing error : unsigned 162\_S1". As she is new, she enquires from her senior {\em Sandra} at every step. {\em Sandra} uses  Cscope~\cite{venkatachalam2005vim}  tool in Vi to grep the code base with the error string and locates  function {\tt VHDLPosedge\#S2} in file {\tt VHDLPosedge.cc} and provides to her. {\em Neha} asks {\em Sandra} for any similar defect. {\em Sandra}  searches ClearQuest, discovers {\tt bug\#22} and searches the  Microsoft Concurrent Versions System (CVS)~\cite{grune1986concurrent} and {\em emails} to extract the bug related commit summary for  code level changes. The summary  stated  a change of data type from {\tt unsigned int} to {\tt unsigned long int} for variable {\tt var1}, but in present code {\tt var1} has type -- {\tt long int} causing {\tt bug\#67}. {\em Sandra} then recalls this change as part of change request {\tt CR123} for optimisation of file {\tt VHDLPosedge.cc} a month ago. {\em Sandra} tells {\em Neha} to revert the  datatype to {\tt unsigned long int}, as it would not affect the behaviour of the code. As part of {\tt CR123},  {\tt VHDLPosedge\#S2} is called by a thread start function, so {\em Sandra}  tells {\em Neha} to add mutex locks for  read  and writes in the function to prevent data-races.

{\em Sandra} responds to  queries of Neha based on multiple relevant sources, and also provides additional important information ({\em Smart} help).  However due to evolving teams and fragmented task distribution, resources like {\em Sandra} who is aware of various aspects of the application, are hardly available. 

\begin{table}[!ht]
	\caption{Knowledge Type -- Knowledge Sources}
	\centering
	\begin{scriptsize}
		\begin{tabular}{|p{4.5cm}|p{3.5cm}|}
			\hline
\multicolumn{1}{|c}{\bf Type, Description and Example}                                                                                                          & \multicolumn{1}{|c|}{\bf Sources}   \\ \hline\hline			
{\bf Software Development}:  domain of programming
like data structure, algorithm, memory, concurrency. Example: {\em variable, data race} &  Source Code, Runtime Trace, Code Comments, Version Trackers (SVN, github, CVS), Bug trackers (JIRA, ClearQuest, Bugzilla), Design Documents, KT sessions \\ \hline	

{\bf Application Oriented}:   entities and actions of the specific application. Example: {\em Convex Hull, bookmark} &   Code Comments, Version Trackers, Bug trackers, Design Documents, KT sessions  \\ \hline	

{\bf Version Evolution}:   commits done for application along with summary. Example: {\em Commit 71\#: Code for new UI functions} &    Version Trackers, Code Comments, KT sessions \\ \hline	

{\bf Defect Evolution}:    bugs reported and fix summary and root cause analysis. Example: {\em BugZilla\#521: Data type mismatch} &    Bug trackers, Code Comments, KT sessions \\ \hline

{\bf Project Management}:    mapping of developer details to defects, version commits, source code elements,  most vulnerable module. Example: {\em Developer A fixed bug BugZilla\#521, developed two modules } &   Code Comments, Version Trackers (SVN, github, CVS), Bug trackers, Emails, KT sessions \\ \hline	

{\bf Business Specs.}:   client \& company details. Example: {\em Business profile of clients} &   Induction manuals, emails, KT sessions  \\ \hline	

		\end{tabular}
		
	\label{tab:knowledge} 
	\end{scriptsize}
\end{table}

We propose a  search framework named {\sc SmartKT} for  single and multi-threaded C / C++ and python code-bases, designed to enact   {\em Sandra} and respond to  common queries of maintenance engineers~\cite{ sillito2008asking, begel2014analyze} related to syntax and semantics of a software. {\sc SmartKT} supports  four types of queries -- a) Entity based~\cite{pound2010ad}:  {\em  Variable D}?, {\em  Function A}? b) List  Search~\cite{pound2010ad}: {\em All static variables of file ftpety.c}, {\em All bugs fixed on 12-03-2013} c) Template based~\cite{pound2010ad}: {\em Which is the algorithm in function \_ \_ \_ \& what are the data-structures used}? {\em Function \_ \_ \_ was effected by which bug numbers \& how many were fixed by Developer \_ \_ \_}? d) Free-form english queries: {\em How many unsynchronised global variables are used to implement the UI Save button}?. For each  query,  {\sc SmartKT} provides direct responses  coupled with {\em Smart} information like {\em responding with additional data race alerts} for  queries  on  {\em global variables}. {\sc SmartKT} additionally helps to bridge missing information across sources and validates  application metadata (like comments).

\begin{figure}[!ht]
	
	\begin{center}
		\includegraphics[scale=0.32]{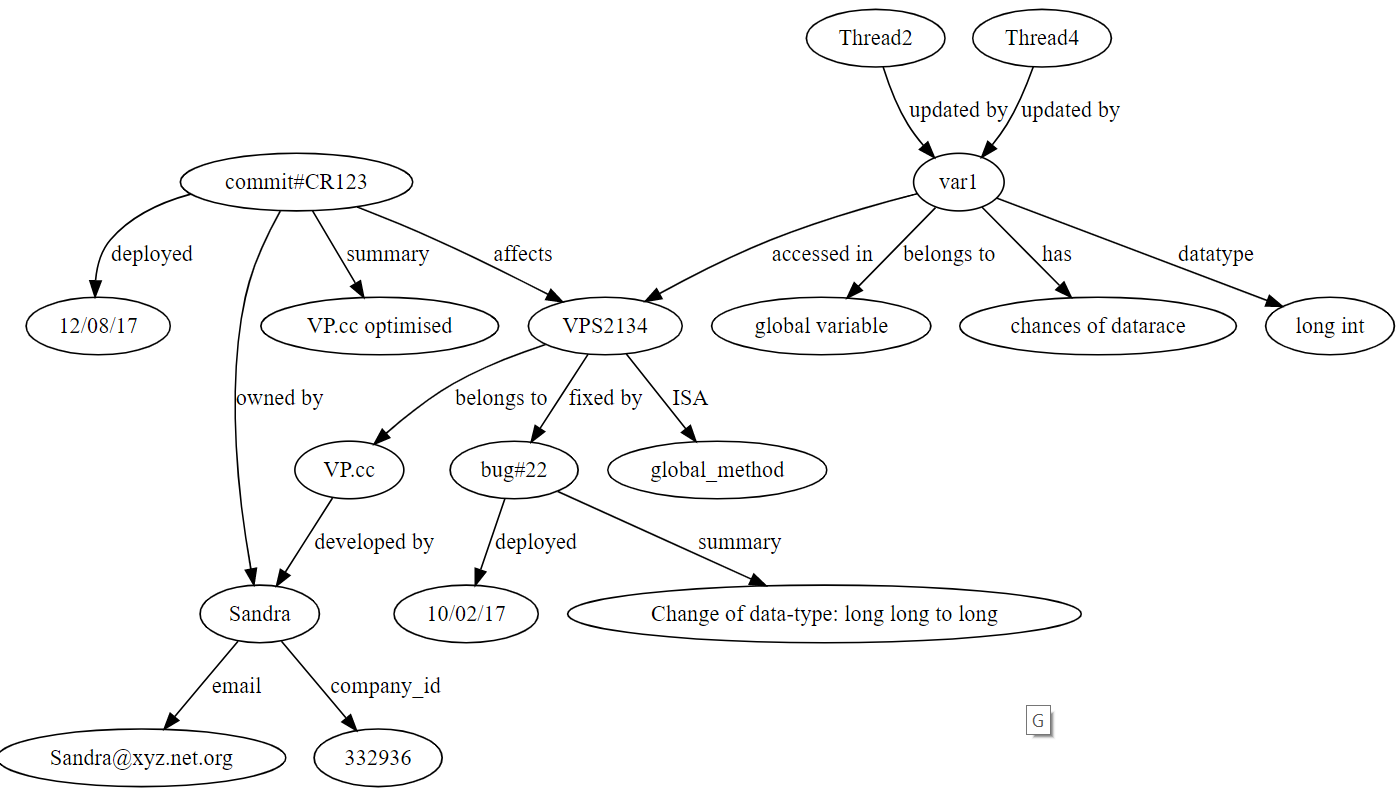} 
		\caption{Knowledge Graph ({\tiny Note: VP stands for VHDLPosedge})}
		\label{fig:graph}
	\end{center}
\end{figure}

\begin{figure}[!ht]
	
	\begin{center}
		\includegraphics[scale=0.30]{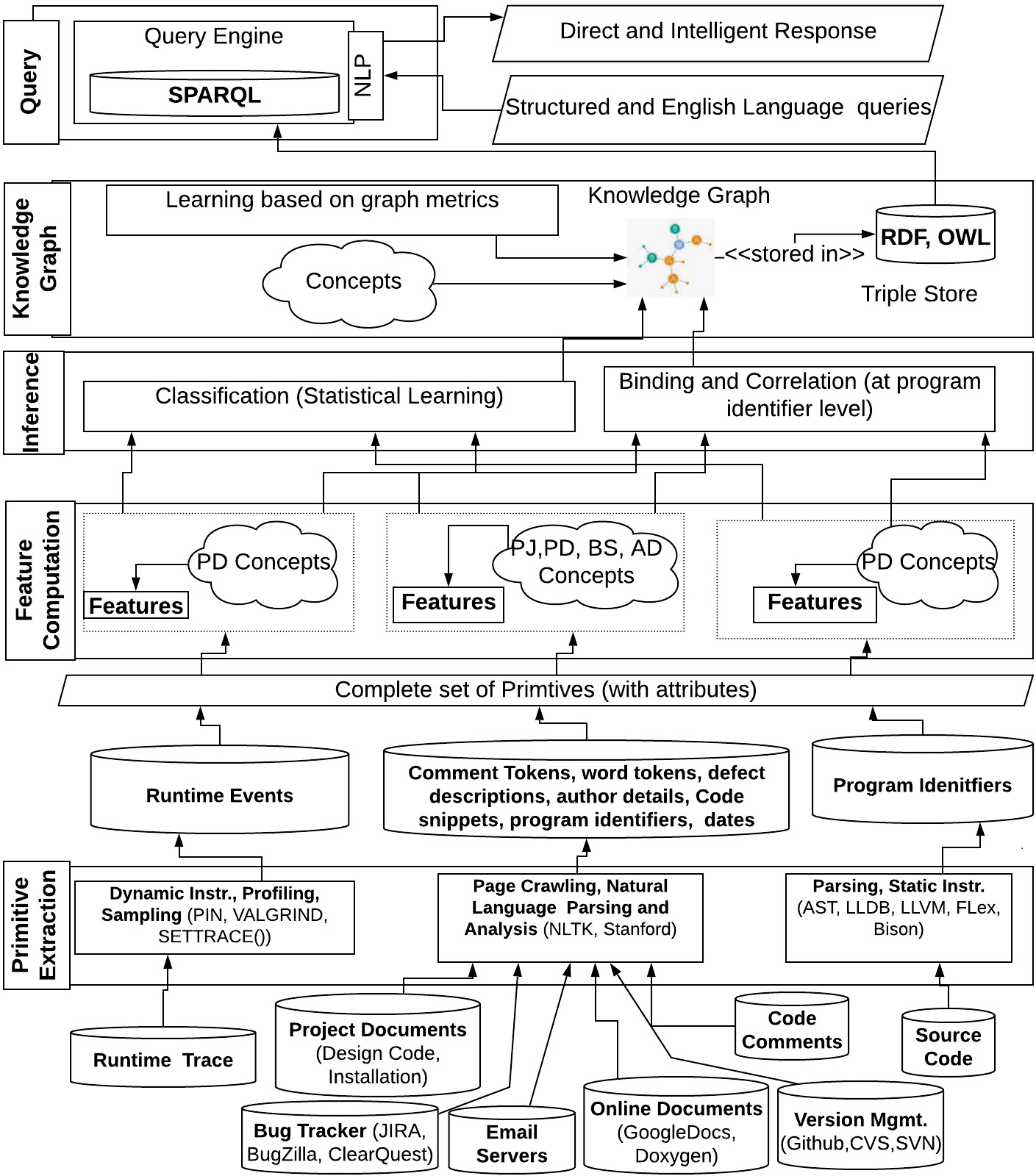} 
		\caption{Architecture of {\sc SmartKT}: Knowledge sources from  Table~\ref{tab:knowledge}}
		\label{fig:archi}
	\end{center}
\end{figure}


 {\sc SmartKT} extracts knowledge from multiple sources (Table~\ref{tab:knowledge}), associates them and represents them in form of a {\em semantic} graph (Figure~\ref{fig:graph} corresponding to {\em Sandra's} knowledge). The design (encompassing graph construction and  query interface) of {\sc SmartKT} (Figure~\ref{fig:archi}) has been incorporated into five layers which we discuss briefly below.

\begin{itemize}
\item {\em Knowledge Primitive Extraction}: In this layer, atomic units of knowledge (referred to as {\em Primitives}) like {\em comment tokens}, {\em bug deployment date}, {\em global data write event}, {\em program identifier} are extracted from the relevant sources using  natural language processing~\cite{chowdhury2003natural} techniques and instrumentation frameworks~\cite{PIN,lattner2004llvm}. 

\item{\em Feature Computation}: In a software,  application specific concepts are modeled in terms of software development concepts (like data-structures, algorithm strategy, concurrency). In this layer, we infer each such concepts from the primitives, using machine learning algorithms. For example, using features extracted from runtime events, structure of the application and  an enumerated ontology for algorithms, we can learn the {\em  classes (greedy, divide and conquer)} for the concept {\em Algorithm Strategy}. 

\item{\em Inference and Knowledge Extraction}: The inferred concepts or knowledge primitives are  mapped to  source code elements to construct knowledge triples (\{{\em primtive/inferred class, association, source code element}\}) and form associations with each other. Example -- {\em Bug\#22}, fixed on {\em 12th July, 2015} (Project Management), affects  {\em Foo1} (source code element) which is a virtual member method (Software Development) of class {\em F}.

\item    {\em Knowledge Graph Construction}: A semantic  graph is constructed based on these triples and stored  in RDF databases~\cite{miller1998introduction}. We  use {\em pagerank}~\cite{xing2004weighted} and {\em triangle counting}~\cite{welzl1988partition} algorithms to learn from  and extend the graph.

\item    {\em Query Processing}: We use SPARQL~\cite{prud2006sparql} to query graphs. For intelligent reponses, we extract related portion of the graphs based on word vector semantics~\cite{levy2014dependency}. For handling English language queries, we plan to employ syntax and semantic matching to a corpus of queries.   
\end{itemize}


A prototype of {\sc SmartKT} to extract software development  and application oriented knowledge with support for entity-based, list and template-based queries has been developed  and is currently under trial by a group of professional developers.



\bibliographystyle{plain}
\bibliography{biblio_thesis}

\begin{thebibliography}{10}

\bibitem{abebe2009analyzing}
Surafel~Lemma Abebe, Sonia Haiduc, Andrian Marcus, Paolo Tonella, and Giuliano
  Antoniol.
\newblock Analyzing the evolution of the source code vocabulary.
\newblock In {\em { European Conference on Software Maintenance and
  Reengineering (ESMR)}}, pages 189--198. IEEE, 2009.

\bibitem{abebe2010natural}
Surafel~Lemma Abebe and Paolo Tonella.
\newblock Natural language parsing of program element names for concept
  extraction.
\newblock In {\em { International Conference on Program Comprehension (ICPC)}},
  pages 156--159. IEEE, 2010.

\bibitem{basthikodi2016classifying}
Mustafa Basthikodi and Waseem Ahmed.
\newblock Classifying a program code for parallel computing against hpcc.
\newblock In {\em { International Conference on Parallel, Distributed and Grid
  Computing (PDGC)}}, pages 512--516. IEEE, 2016.

\bibitem{begel2014analyze}
Andrew Begel and Thomas Zimmermann.
\newblock Analyze this! 145 questions for data scientists in software
  engineering.
\newblock In {\em { International Conference on Software Engineering (ICSE)}},
  pages 12--23. ACM, 2014.

\bibitem{chatterjee2015debugging}
Nachiketa Chatterjee, Srijoni Majumdar, Shila~Rani Sahoo, and Partha~Pratim
  Das.
\newblock Debugging multi-threaded applications using pin-augmented gdb (pgdb).
\newblock In {\em { International Conference on Software Engineering Research
  and Practice (SERP)}}, pages 109--115. WorldComp-USA, 2015.

\bibitem{chowdhury2003natural}
Gobinda~G Chowdhury.
\newblock Natural language processing.
\newblock {\em Annual review of information science and technology, Wiley},
  37(1):51--89, 2003.

\bibitem{vcubranic2003hipikat}
Davor {\v{C}}ubrani{c} and Gail~C Murphy.
\newblock Hipikat: Recommending pertinent software development artifacts.
\newblock In {\em { International Conference on Software Engineering (ICSE)}},
  pages 408--418. ACM, 2003.

\bibitem{prud2006sparql}
Richard Cyganiak.
\newblock A relational algebra for sparql.
\newblock {\em Digital Media Systems, HP Laboratories}, 35:9, 2005.

\bibitem{dehaghani2013factors}
Sayed Mehdi~Hejazi Dehaghani and Nafiseh Hajrahimi.
\newblock Which factors affect software projects maintenance cost more?
\newblock {\em Acta Informatica Medica, The Academy of Medical Sciences of
  Bosnia and Herzegovina}, 21(1):63--72, 2013.

\bibitem{erlikh2000leveraging}
Len Erlikh.
\newblock Leveraging legacy system dollars for e-business.
\newblock {\em IT professional, IEEE}, 2(3):17--23, 2000.

\bibitem{fluri2007code}
Beat Fluri, Michael Wursch, and Harald~C Gall.
\newblock Do code and comments co-evolve? on the relation between source code
  and comment changes.
\newblock In {\em { Working Conference on Reverse Engineering (WCRE)}}, pages
  70--79. IEEE, 2007.

\bibitem{freitas2008role}
Jos{e}~Lu{\i}s Freitas, Daniela da~Cruz, and Pedro~Rangel Henriques.
\newblock The role of concepts on program comprehension.
\newblock In {\em { International Conference on Program Comprehension (ICPC)}}.
  ACM, 2011.

\bibitem{grune1986concurrent}
Dick Grune et~al.
\newblock {\em Concurrent versions systems, a method for independent
  cooperation}.
\newblock VU Amsterdam. Subfaculteit Wiskunde en Informatica, 1986.

\bibitem{ko2006exploratory}
Andrew~J Ko, Brad~A Myers, Michael~J Coblenz, and Htet~Htet Aung.
\newblock An exploratory study of how developers seek, relate, and collect
  relevant information during software maintenance tasks.
\newblock {\em { IEEE Transactions on software engineering}, IEEE},
  32(12):971--987, 2006.

\bibitem{koskinen2003software}
Jussi Koskinen.
\newblock Software maintenance costs.
\newblock Technical report, Information Technology Research Institute,
  University of Jyvaskyla, 2003.

\bibitem{lamb1998learning}
Linda Lamb, Arnold Robbins, and Arthur Robbins.
\newblock {\em Learning the vi Editor}.
\newblock " O'Reilly Media, Inc.", 1998.

\bibitem{lattner2004llvm}
Chris Lattner and Vikram Adve.
\newblock The llvm compiler framework and infrastructure tutorial.
\newblock In {\em { International Workshop on Languages and Compilers for
  Parallel Computing (LCPC)}}, pages 15--16. Springer, 2004.

\bibitem{levy2014dependency}
Omer Levy and Yoav Goldberg.
\newblock Dependency-based word embeddings.
\newblock In {\em { Annual Meeting of the Association for Computational
  Linguistics (ACL)}}, pages 302--308. ACM, 2014.

\bibitem{PIN}
Chi-Keung Luk, Robert Cohn, Robert Muth, Harish Patil, Geoff Lowney, Steven
  Wallace, Vijay~Janapa Reddi, and Kim Hazelwood.
\newblock Pin: building customized program analysis tools with dynamic
  instrumentation.
\newblock In {\em { Special Interest Group on programming languages notices
  (SIGPLAN)}}, pages 190--200. ACM, 2005.

\bibitem{majumdar2022automated}
Srijoni Majumdar, Ayush Bansal, Partha~Pratim Das, Paul~D Clough, Kausik Datta,
  and Soumya~Kanti Ghosh.
\newblock Automated evaluation of comments to aid software maintenance.
\newblock {\em Journal of Software: Evolution and Process}, 34(7):e2463, 2022.

\bibitem{majumdar2021mathematical}
Srijoni Majumdar, Nachiketa Chatterjee, Partha~Pratim Das, and Amlan
  Chakrabarti.
\newblock A mathematical framework for design discovery from multi-threaded
  applications using neural sequence solvers.
\newblock {\em Innovations in Systems and Software Engineering},
  17(3):289--307, 2021.

\bibitem{majumdar2021dcube_}
Srijoni Majumdar, Nachiketa Chatterjee, Partha Pratim~Das, and Amlan
  Chakrabarti.
\newblock Dcube\_ nn d cube nn: Tool for dynamic design discovery from
  multi-threaded applications using neural sequence models.
\newblock {\em Advanced Computing and Systems for Security: Volume 14}, pages
  75--92, 2021.

\bibitem{majumdar2016d}
Srijoni Majumdar, Nachiketa Chatterjee, Shila~Rani Sahoo, and Partha~Pratim
  Das.
\newblock D-cube: Tool for dynamic design discovery from multi-threaded
  applications using pin.
\newblock In {\em { International Conference on Software Quality, Reliability
  and Security (QRS)}}, pages 25--32. IEEE, 2016.

\bibitem{majumdar2020comment}
Srijoni Majumdar, Shakti Papdeja, Partha~Pratim Das, and Soumya~Kanti Ghosh.
\newblock Comment-mine—a semantic search approach to program comprehension
  from code comments.
\newblock In {\em Advanced Computing and Systems for Security}, pages 29--42.
  Springer, 2020.

\bibitem{majumdar2019s}
Srijoni Majumdar, Papdeja Shakti, Partha~Pratim Das, and Soumya Ghosh.
\newblock Smartkt: A search framework to assist program comprehension using
  smart knowledge transfer.
\newblock In {\em { International Conference on Software Quality, Reliability
  and Security (QRS)}}, pages 97--108. IEEE, 2019.

\bibitem{miller1998introduction}
Eric Miller.
\newblock An introduction to the resource description framework.
\newblock {\em Bulletin of the American Society for Information Science and
  Technology, Wiley}, 25(1):15--19, 1998.

\bibitem{pigoski1996practical}
Thomas~M Pigoski.
\newblock {\em Practical software maintenance: best practices for managing your
  software investment}.
\newblock Wiley, 1996.

\bibitem{pound2010ad}
Jeffrey Pound, Peter Mika, and Hugo Zaragoza.
\newblock Ad-hoc object retrieval in the web of data.
\newblock In {\em { International conference on World Wide Web}}, pages
  771--780. ACM, 2010.

\bibitem{razzaq2018study}
Seher Razzaq, Yan-Fu Li, Chu-Ti Lin, and Min Xie.
\newblock A study of the extraction of bug judgment and correction times from
  open source software bug logs.
\newblock In {\em { International Conference on Software Quality, Reliability
  and Security Companion (QRS)}}, pages 229--234. IEEE, 2018.

\bibitem{roehm2012professional}
Tobias Roehm, Rebecca Tiarks, Rainer Koschke, and Walid Maalej.
\newblock How do professional developers comprehend software?
\newblock In {\em { International Conference on Software Engineering (ICSE)}},
  pages 255--265. IEEE, 2012.

\bibitem{shepherd2012sando}
David Shepherd, Kostadin Damevski, Bartosz Ropski, and Thomas Fritz.
\newblock Sando: an extensible local code search framework.
\newblock In {\em { International Symposium on the Foundations of Software
  Engineering (FSE)}}, page~15. ACM, 2012.

\bibitem{sillito2008asking}
Jonathan Sillito, Gail~C Murphy, and Kris De~Volder.
\newblock Asking and answering questions during a programming change task.
\newblock {\em { IEEE Transactions on Software Engineering}, IEEE},
  34(4):434--451, 2008.

\bibitem{trumper2010understanding}
Jonas Trumper, Johannes Bohnet, and Jurgen Dollner.
\newblock Understanding complex multithreaded software systems by using trace
  visualization.
\newblock In {\em { International Symposium on Software Visualization (ISSV)}},
  pages 133--142. ACM, 2010.

\bibitem{varanda2012problem}
Maria~Jo{\~a}o Varanda~Pereira, Mario~Marcelo Beron, Daniela da~Cruz, Nuno
  Oliveira, and Pedro~Rangel Henriques.
\newblock Problem domain oriented approach for program comprehension.
\newblock In {\em { OpenAccess Series in Informatics (OASIcs)}}, pages 21--37.
  Schloss Dagstuhl-Leibniz-Zentrum fuer Informatik, 2012.

\bibitem{venkatachalam2005vim}
Girish Venkatachalam.
\newblock Vim for c programmers.
\newblock {\em Linux Journal, Slashdot Group}, 2005(140):9, 2005.

\bibitem{wahlisoftware}
Ueli Wahli et~al.
\newblock Software configuration management a clear case for ibm rational
  clearcase and clearquest ucm, dec. 2004.
\newblock {\em IBM,}.

\bibitem{welzl1988partition}
Emo Welzl.
\newblock Partition trees for triangle counting and other range searching
  problems.
\newblock In {\em { Annual symposium on Computational geometry}}, pages 23--33.
  ACM, 1988.

\bibitem{xing2004weighted}
Wenpu Xing and Ali Ghorbani.
\newblock Weighted pagerank algorithm.
\newblock In {\em { Communication Networks and Services Research}}, pages
  305--314. IEEE, 2004.

\end{thebibliography}
\end{document}